\documentstyle[12pt,aasms4]{article}

\def\etal{{et al.}}
\def\asca{{\it ASCA}}

\def\gi{{\it Ginga}}

\def\delchi{{$\Delta \chi^{2}$}}

\def\Msun{\hbox{$\rm\thinspace M_{\odot}$}}

\begin{document}

\title{On the dependence of the iron K-line profiles with luminosity
in Active Galactic Nuclei}

\author {K. Nandra\altaffilmark{1,2}, 
I.M. George \altaffilmark{1, 3}, R.F. Mushotzky\altaffilmark{1}, 
T.J. Turner \altaffilmark{1, 3}, T. Yaqoob\altaffilmark{1, 3}}

\altaffiltext{1}{Laboratory for High Energy Astrophysics, Code 660,
	NASA/Goddard Space Flight Center,
  	Greenbelt, MD 20771}
\altaffiltext{2}{NAS/NRC Research Associate}
\altaffiltext{3}{Universities Space Research Association}

\slugcomment{Submitted to {\em The Astrophysical Journal Letters}}

\begin{abstract}

We present evidence for changes in the strength and profile 
of the iron K$\alpha$ line in Active Galactic Nuclei (AGN), based
on X-ray observations with \asca. There is a clear decrease
in the strength of the line with increasing luminosity. This relation is
is not due solely to radio power, as it persists when only
radio-quiet AGN are considered and therefore cannot be
fully explained by relativistic beaming. In addition to the
change in strength, the line profile also appears to be
different in higher luminosity sources. We discuss these results
in terms of a model where the accretion disk becomes ionized as
a function of the accretion rate.
\end{abstract}


\section{Introduction}
\label{sec:intro}

Seyfert 1 galaxies exhibit iron K$\alpha$ emission lines in their
X-ray spectra which are characteristic of relativistic effects in an
accretion disk surrounding a central black hole (Tanaka \etal\ 1995;
Yaqoob \etal\ 1995; Nandra \etal\ 1997 hereafter N97). These lines can
be used as a diagnostic of the innermost regions of AGN, and therefore
merit further study in classes other than Seyfert 1s. The iron K$\alpha$
emission was first studied in detail using the \gi\ spectra of Seyfert
galaxies (Nandra \& Pounds 1994 and references therein) and based on
these results, Iwasawa \& Taniguchi (1993, hereafter IT93) suggested
that there may be an X-ray ``Baldwin Effect'' whereby the equivalent
width (EW) of the emission line reduced with increasing
luminosity. However, this result has been disputed (Nandra \& Pounds
1994) and it was unclear whether the correlation held when emission
lines from ``quasars'' were considered (Williams \etal\ 1992;
IT93). With far greater sensitivity and spectral resolution than \gi,
\asca\ (Tanaka, Inoue \& Holt 1994) can be used to provide a more
stringent test of such an hypothesis. \asca\ results for individual
sources have been suggestive that this trend might hold (e.g., Elvis
\etal\ 1994; Nandra \etal\ 1995). However, a systematic comparison
requires consideration of a larger number of sources.  The results of
an analysis of the dependence of iron line properties with luminosity,
based on a sample of 39 AGN with broad optical lines, will be 
the subject of this
{\it Letter}.

\section{Results}
\label{sec:results}

Our intention is to investigate the iron K$\alpha$ properties,
primarily as a function of luminosity, although other relevant
parameters are also considered. N97 have investigated the iron line
properties of 18 Seyfert 1 galaxies in detail, and we have used those
data here. To extend the range of luminosities, we also include iron
line data from a sample of 21 quasars presented by Nandra \etal\ 
(1998; hereafter N98). ``Seyfert 1 galaxies'' are defined by N97 as AGN with
predominantly broad optical lines at redshift $z<0.05$ and ``QSOs''
defined by N98 as broad-line AGN at $z>0.05$. Neither sample is
well-selected in the conventional sense, and we will discuss this
below.  Further details of the analysis of the observations can be
found in N97 and N98.  In both classes of source, the continuum is
well approximated by a power-law in the 3-10 keV band and we use such
a parameterization here. In that band, we expect little contamination
from other components, with the only major concern being the presence
of an iron K-absorption edge from the ``Compton reflection'' continuum
which accompanies the iron line (e.g., George \& Fabian
1991). However, that edge is very weak when contrasted against the
direct continuum and has little effect on the inferred properties of
the iron line (N97). As the signal-to-noise ratio for many of the
individual objects, and particularly the quasars, can be rather low,
we have assembled mean line profiles for the whole sample and various
subsamples for comparison. Following N97, we constructed these by
fitting a power-law to the \asca\ SIS spectra in the 3-10 keV band,
excluding the ``iron band'' from 5-7 keV (all energies 
are quoted in the rest frame).
Such an interpolation allows us to investigate the iron line
properties without excessive model dependence. The data/model ratios
for each source were transformed into the rest frame in each case,
which then permits us to co-add the residuals to produce the mean line
profiles.

\subsection{Profiles as a function of AGN class}

\begin{figure}
\epsscale{0.9}
\plotone{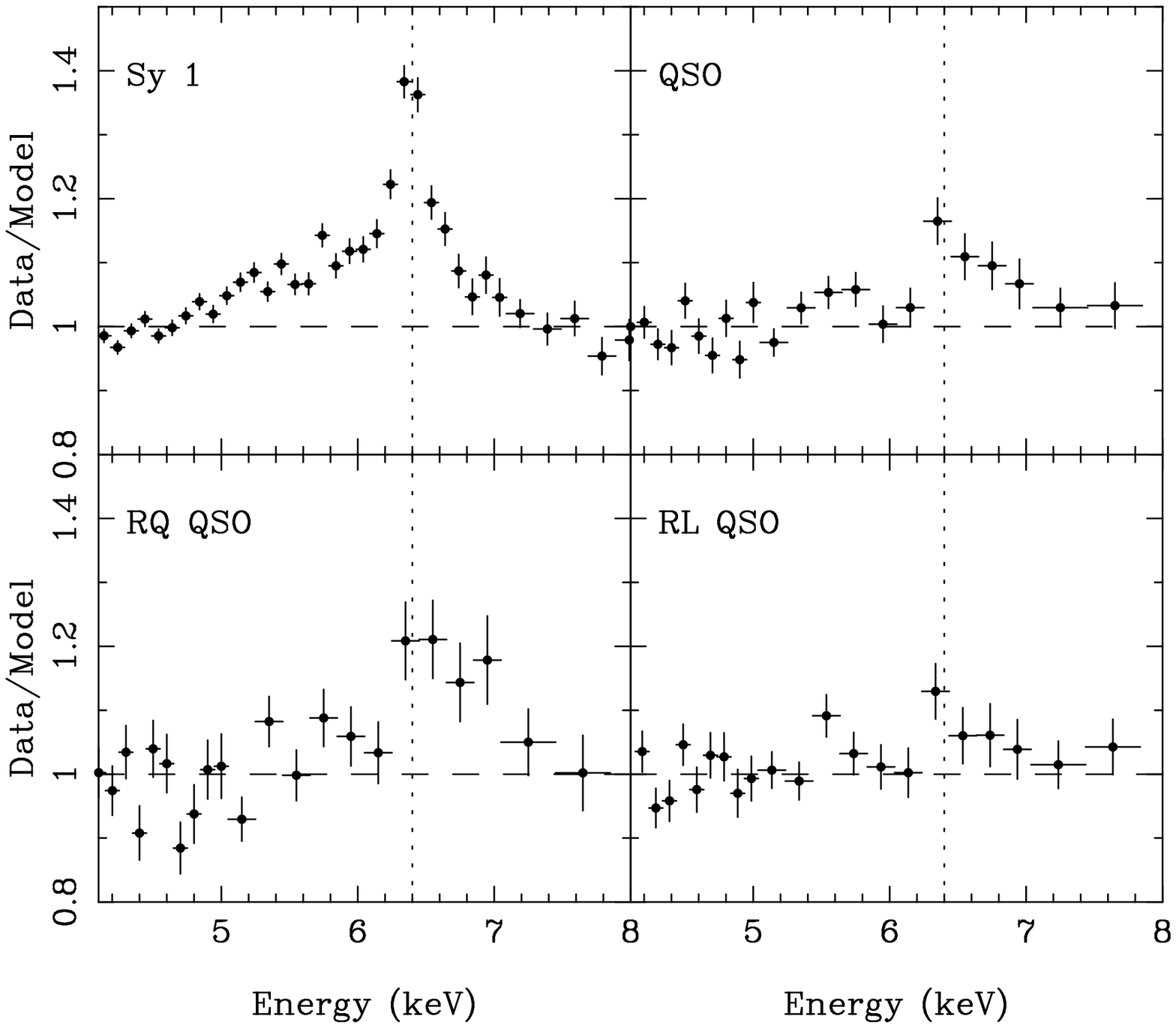}
\caption{
Mean data/model ratios, transformed into the rest frame, assuming a
power-law fit to the 3-10~keV band in the rest frame, excluding the
5-7 keV ``iron band''. The upper left panels shows the Seyfert 1
profile (N97). The upper right panel shows the QSOs of N98. The lower
left panel shows the ``radio quiet'' sources from the N98 sample and
the lower right the ``radio-loud'' sources (see text).  The vertical
dotted lines are at an energy of 6.4 keV. There is a clear difference
between the Seyfert 1s and the QSOs, with the latter showing a weaker
line, a weaker ``red wing'' and relatively higher ``blue'' flux. When
radio quiet QSOs are considered alone, differences with the Seyfert 1s
remain, with at least the core being significantly different. Radio
loud sources exhibit little line emission at all, possibly because
this subsample is dominated by very high luminosity objects.
\label{fig:profs_class}} 
\end{figure}

For initial comparison, we show in Fig~\ref{fig:profs_class}
the mean ratios for the Seyfert 1 galaxies and the QSOs. This splits
the sample into low ($z<0.05$) and high ($z>0.05$) redshift
objects. There is a very clear difference between the two, with the
QSOs showing much weaker emission, little evidence for the strong
``red wing'' characteristic of the gravitational effects of the black
hole, and relatively stronger ``blue'' flux (i.e. emission above the
rest energy of neutral iron at 6.4~keV). It has been proposed that the
weakness of the emission lines in some high-redshift AGN may be due to
the fact that the X-ray emission is produced in a relativistic jet
which is beamed away from the accretion disk. However, it is highly
unlikely that this is the sole reason for the difference in the
character of the lines in the Seyfert and QSO samples.  To demonstrate
this we show in Fig~\ref{fig:profs_class}c and d the line profiles for
``radio quiet'' and ``radio loud'' subsamples.  ``Radio loud'' sources
are defined by N98 as those having $\log (f_{\rm 5 Ghz}/f_{\rm V}) >
1$, where $f_{\rm 5 Ghz}$ is the ratio of the flux at 5 GHz and $f_{\rm V}$
the optical flux in the V-band. Radio-loud quasars are often separated
from radio-quiet objects, as they contain relativistic jets. It can be
seen from these panels that the radio-loud subsample does indeed show
weaker line emission. But even when only radio-quiet AGN are
considered there is a clear difference between the profiles of AGN at
$z<0.05$ and those which are more distant (note that the N97 sample
contains only one radio loud source, and is thus dominated by
radio-quiet objects). Another parameter, the X-ray luminosity, is
highly correlated with both redshift and radio-loudness in the QSO
(and also the combined) sample. We now investigate the dependence of
the iron line properties with source luminosity.

\subsection{Dependence on luminosity}

Fig.~\ref{fig:profs_lumin} shows the line profiles split into 5 bins
based on the mean X-ray luminosity of the source in the 2-10 keV band,
$L_{\rm X}<10^{43}$~erg s$^{-1}$ (6 observations of 4 sources, no radio loud
source), $10^{43}<L_{\rm X}<10^{44}$~erg s$^{-1}$ (15 observations of 11
sources, none radio-loud), $10^{44}<L_{\rm X}<10^{45}$~erg s$^{-1}$ (9
observations of 9 sources, 3 radio loud), $10^{45}<L_{\rm
X}<10^{46}$~erg s$^{-1}$ (4 observations of 4 sources, 1 radio loud),
$L_{\rm X}>10^{46}$~erg s$^{-1}$ (11 observations of 11 sources, 9 radio
loud). Clearly then, the vast majority of the radio-loud AGN are
present in this highest-luminosity bin.  Also shown is the mean
profile of the combined Seyfert/QSO sample. This latter plot is
dominated by the high signal-to-noise observations of the Seyfert
galaxies, which can be seen by comparing that profile to those of the
two lowest luminosity bins, which consist entirely of Seyferts. These
three profiles are remarkably similar.

However, there is a clear change in the profile for luminosities
$L_{\rm X}>10^{44}$~erg s$^{-1}$.  The bin with $10^{44}<L_{\rm
X}<10^{45}$~erg s$^{-1}$ shows a weakening of the line emission,
particularly at the core and relatively stronger blue flux, but still
with evidence for the redshifted wing. At $10^{45}<L_{\rm
X}<10^{46}$~erg s$^{-1}$ we see no longer see any evidence for any red
wing, and the peak line flux occurs at an energy higher than
6.4~keV. A gaussian fit to the profile in this bin shows a best-fit
energy of 6.57~keV, and is inconsistent with 6.4~keV at $>99$~per cent
confidence (\delchi=12.5). The line has weakened further, although we
caution that there are few objects in this luminosity bin. Above
$L_{\rm X}=10^{46}$~erg s$^{-1}$, there is no evidence for {\it any}
line emission and the level of any undetected emission is clearly well
below any of the other profiles. Our data therefore strongly imply
that the strength of the iron K$\alpha$ lines in AGN reduces as a
function of increasing luminosity confirming the X-ray ``Baldwin''
effect of IT93. Furthermore, we see good evidence for changes in the
profile of the line with luminosity, with high-luminosity sources
showing a weaker core and red wing, and stronger blue flux up to the
point where the line emission disappears altogether. We quantify the
dependence of line EW on luminosity in Fig.~\ref{fig:lum_ew}, which
shows the mean EW of the emission lines as a function of X-ray
luminosity for the bins described above.  Two measures are presented:
an estimate of the ``core'' EW, which has been modeled as a narrow
gaussian, and the ``total'' EW, which is modeled as a relativistic
disk-line in the case of the Seyferts, and a broad gaussian in the
case of the quasars (N97, N98).  There is a strong anti-correlation
between the luminosity and EW in both cases.  The total EW in
particular shows a strong reduction above $L_{\rm x}=10^{45}$~erg
s$^{-1}$. In the two highest-luminosity bins the core and total EWs
are consistent, indicating that the broad wings of the emission line
have disappeared (Fig.~\ref{fig:profs_lumin}).

\subsection{Selection Effects}

Our samples may suffer from substantial selection effects and biases
due to correlated parameters, as intimated above. The objects
are selected from the \asca\ public archive, and presence in that archive
requires no clearly defined scientific criteria. 
However, the tendency is for the sources to be
X-ray selected. Bright sources tend to be observed early in any X-ray
mission and thus enter the archives first. Our sample also suffers
from a bizarre redshift bias due to selection by the time allocation
committees, whereby we are dominated by very low redshift objects
($z<0.1$) which are X-ray bright and very high redshift sources
($z>1$) which provide potentially exciting results.  The very highest
luminosity bin is dominated by radio-loud sources some of which are at
very high redshift (z=3-4). Therefore, our changes in profile with luminosity
could be attributable to changes with radio-loudness, or redshift. As
discussed above, the former cannot fully explain the observed changes,
as we do find strong line emission in low-luminosity, radio-loud
AGN. Similarly, cosmological epoch seems unlikely to be the sole
factor in determining the changes in profile.  Below $10^{45}$~erg
s$^{-1}$, where clear changes in strength and profile are already
occuring, the highest-redshift source has $z=0.129$.  Thus
redshift-evolution would have to occur extremely rapidly to explain
our results.  On the other hand a luminosity effect could be the {\it
sole} source of our observed correlation, there being no evidence to
the contrary.  Above a luminosity of $10^{45}$~erg s$^{-1}$, only 2/13
sources show evidence for line emission at all, E1821+643 and MR
2251-178. In the latter case the evidence is rather marginal (N98). In
the former there still remains some confusion as to whether some
fraction of the line arises from the surrounding cluster, but the
emission is probably dominated by the QSO (Kii \etal\ 1991; Yamashita
\etal\ 1997). On the other hand, all but one (PG 1404+226) of the 24
sources with luminosity below $10^{45}$~erg s$^{-1}$ show evidence for
line emission. The lack of a detection in the case of PG 1404+226 can
be attributed to the low signal-to-noise ratio of the spectrum in the
hard X-ray band.  Another point to consider is that radio-loud QSOs,
which therefore have high-luminosity and high-redshift in our
sample, appear to have flatter continuum slopes in the soft X-ray band
(Wilkes \& Elvis 1987). It is unclear whether or not this effect
occurs in the hard X-rays (William \etal\ 1992; Lawson \etal\ 1992;
Lawson \& Turner 1997).  Different continua can affect the strength of
the iron line in that for flatter slopes, higher EWs are
produced (e.g., George \& Fabian 1991).  We would thus expect {\it
stronger} emission lines in the radio-loud sources if they had flatter
hard X-ray slopes, rather than weaker ones, as we observe. We
therefore believe that differences in continuum shape cannot account
for our results.

\section{Discussion}
\label{sec:discuss}

Using a (poorly-selected) sample of X-ray observations of broad-line
AGN, we have shown clear evidence of an X-ray ``Baldwin'' effect,
i.e. a reduction in the strength of the iron K$\alpha$ line with
increasing luminosity. Such an effect was originally suggested based
on \gi\ data by IT93. Although the effect could be partially due to
beaming of the X-rays away from the putative accretion disk in
radio-loud sources, there is still strong evidence for differences
between high and low luminosity sources when only radio-quiet objects
are considered. We therefore conclude that the primary effect is most
likely with source luminosity and discuss the impact of our results in
that context. The emission line in low-luminosity AGN is thought to
arise from an accretion disk, where Doppler and gravitational effects
produce the extreme broadening, and especially the red wing. An
alternative origin for a line core at 6.4~keV is in the putative
molecular torus which may obscure the line-of-sight to Seyfert 2
galaxies (Ghisselini, Haardt \& Matt 1994; Krolik, Madau \& Zycki
1994). In principle then, differences in the line profiles could arise
from differing contributions from the accretion disk and torus. We
observe an effect consistent with this, in that the 6.4~keV peak
reduces with increasing luminosity and largely disappearing above
$L_{\rm X} > 10^{45}$~erg s$^{-1}$ (Fig.~\ref{fig:lum_ew}). The upper
limit to the EW of any narrow, 6.4 keV line in the highest luminosity
bin is $\sim 25$~eV, so our data are consistent with a small
contribution from the torus in all sources, and if that contribution
decreased when the luminosity increased it would account for some of
the differences in line profiles. However, we also observe an effect
that the red wing reduces with luminosity, implying that the disk-line
component also changes.  Indeed, the entire effect can be attributed
to changes in the disk-line.

\begin{figure}
\epsscale{1.0}
\epsfysize=0.85\textwidth
\plotone{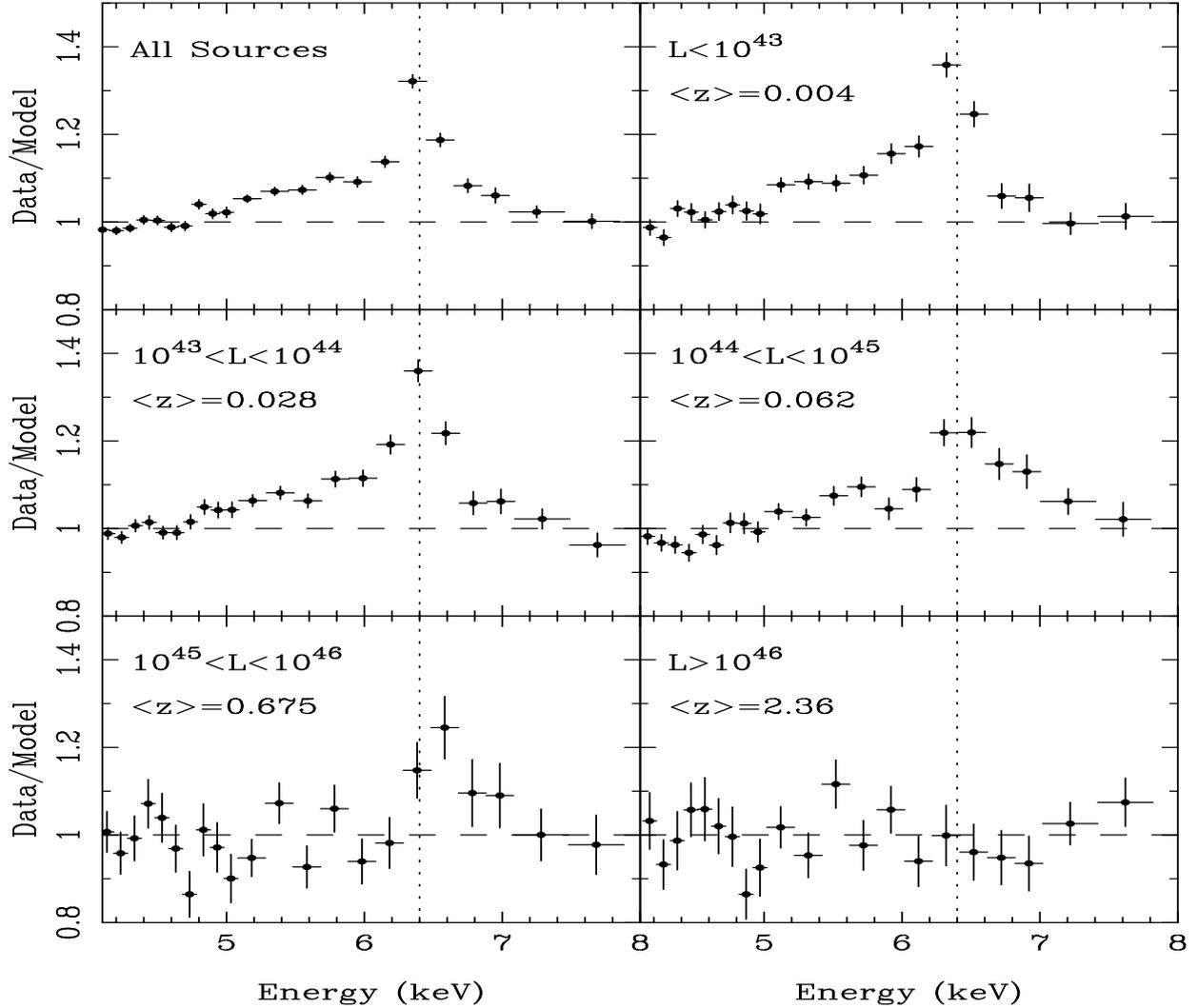}
\caption{
Data/Model ratios, constructed as in Fig.~\ref{fig:profs_class}, split
into luminosity bins.  The vertical dotted lines are at an energy of
6.4 keV.  The profile for all sources (upper left) is dominated by the
high signal-to-noise objects which are mostly low luminosity. Below
$L_{\rm X}=10^{44}$ the line profiles are very similar, but above this
luminosity there are clear changes. The line strength reduces with
increasing luminosity, in both the core and red wing, but the blue
flux is enhanced relative to the total line emission. Above $L_{\rm
X}=10^{46}$ no evidence for line emission is observed at all.  This
confirms the ``X-ray Baldwin'' effect suggested by IT93. The mean
redshift is also shown, and demonstrates the strong correlation
between redshift and luminosity in our sample.
\label{fig:profs_lumin}}
\end{figure}

\begin{figure}
\epsscale{1.0}
\plotone{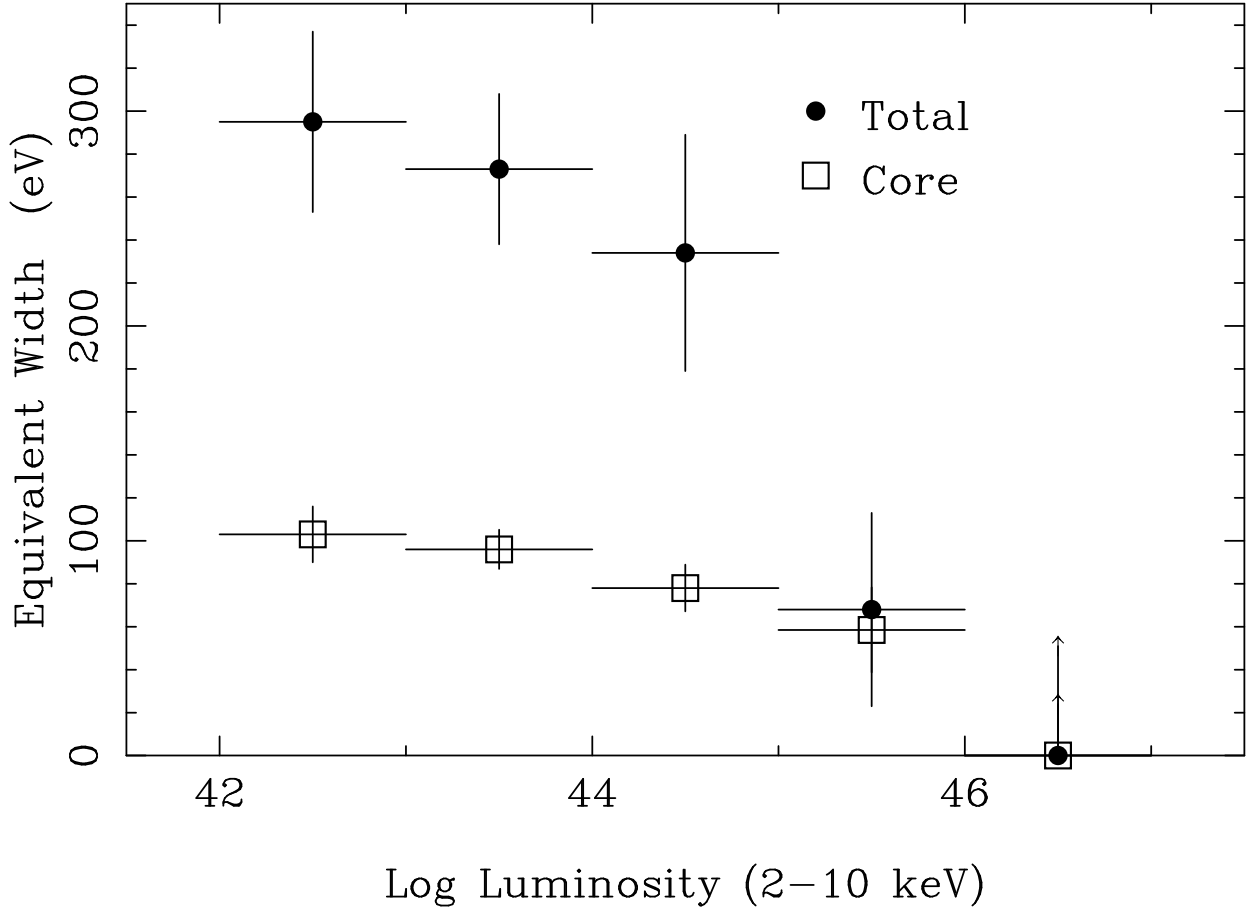}
\caption{
Equivalent width of the iron K$\alpha$ emission lines (rest frame)
versus luminosity, binned as for Fig.~\ref{fig:profs_lumin}. The open
squares show the results for a ``narrow'' gaussian fit, with upper
limits derived for a line at 6.4~keV. These fits tend to model the
line core.  The solid circles are the equivalent widths for a
relativistic disk line in the case of the Seyfert galaxies (N97), and
a broad gaussian with free width for the QSOs in which a line is
detected (N98). Where no line is detected upper limits were determined
for a line at 6.4 keV, with a fixed width of $\sigma=0.43$~keV, the
mean value for Seyferts (N97). These fits will tend to model the total
flux of the emission line better than a narrow gaussian, as evidenced
by the fact that the EWs are higher. Both the core and total EWs show
a clear decrease EW with luminosity. We also note that
above $L_{\rm X} = 10^{45}$~erg s$^{-1}$, the majority of the flux is
modeled by a narrow line, consistent with the lack of a ``red wing''
in the data.
\label{fig:lum_ew}}
\end{figure}

Nandra \etal\ (1995) suggested that the lack of significant iron line
emission in high luminosity AGN might be due to the fact that those
sources have a high accretion rate, causing the disk to become ionized
(Matt, Fabian \& Ross 1993), with iron being fully stripped.  Some
support from that hypothesis came with the detection of an emission
line consistent with highly-ionized iron in an ``intermediate''
luminosity QSO, PG 1116+215 (Nandra \etal\ 1996). Our results are
interpretable in this context. For a given black hole mass, higher
luminosity sources should have a higher accretion rate, as well as
more intense X-ray (ionizing) luminosity. Both would tend to strip
atoms in the disk. At some point, iron will begin to be ionized, which
should cause more ``blue'' flux to be observed from high-ionization
species. In these intermediate ionization states, resonance scattering
can also cause a reduction in the line flux (Matt, Fabian \& Ross
1993, 1996). An increase in the effective fluorescence yield in the
He-like and H-like states would cause stronger line emission when
those species are dominant, but if the emission comes from a range of
radii (and therefore ionization state) in the disk, which appears to
be the case (N97), that effect may not be clearly observable.  At
another transition point, iron atoms in the inner disk will begin to
become fully stripped, which would cause a reduction in the ``red
wing'' and a shift of the mean energy above 6.4~keV. When iron becomes
fully stripped throughout the X-ray illuminated part of the disk, no
emission line will be observed from the disk at all. All of these
effects are observed in Fig.~\ref{fig:profs_lumin}. If this model is
correct, we should observe associated changes in the Compton
reflection component (e.g., Zycki \& Czerny 1994). As the ionization
rises, the disk becomes more reflective in the soft X-ray band,
causing a ``soft excess'', with the potential for associated line
emission from elements lighter than iron (e.g. O, Ne). Again, for very
high ionization states, we see the Compton reflection without it
suffering absorption in the disk, making the ``contrast'' with the
continuum very low, resulting in an apparently-weak Compton hump. Such
effects are only easily testable with instruments with better
high-energy efficiency than \asca.

Of course there may be significant differences in the black hole mass
when moving from low to high luminosity sources. Estimates of the
black hole masses of local AGN have been made based on various
arguments such as optical/UV line widths, stellar kinematics and maser
observations (e.g., Koratkar \& Gaskell 1991; Ford \etal\ 1994;
Miyoshi \etal\ 1995).  Interestingly, these masses tend to lie in the
region $10^{7-8}$~\Msun, regardless of the technique employed. Larger
masses would be expected for the highest-luminosity objects in our
sample, since if their emission is isotropic, their luminosities would
exceed the Eddington limit unless $M>10^{9}$~\Msun. However, it is
interesting to note that for a $10^{8}$~\Msun\ hole, the transition to
$\sim 10$~per cent Eddington accretion occurs at $10^{45}$~erg
s$^{-1}$. This is approximately where Matt \etal\ (1993) predict that
the disk will start to become significantly ionized. and where we
begin to see a change in the line profile.  Super-Eddington accretion
occurs at $10^{46}$~erg $s^{-1}$, above which luminosity the emission
line disappears. Thus we speculate that the AGN in our sample cover a
relatively small mass range, and that the differences in the source
luminosities are due to differences in accretion rate, which then
affect the line profiles. The implication of this is that the quasar
phenomenon is short-lived.

Further exploration of these models awaits the formation of larger,
and preferably complete, well-selected samples within the \asca\
archive, which we anticipate within the next few years.

\acknowledgements 
{We are grateful to the
\asca\ team for their operation of the satellite, the \asca\ GOF at
NASA/GSFC for their assistance in data analysis
This research has made use of the Simbad database, operated
at CDS, Strasbourg, France; of the NASA/IPAC Extragalactic database,
which is operated by the Jet Propulsion Laboratory, Caltech, under
contract with NASA; and data obtained through the HEASARC on-line
service, provided by NASA/GSFC. We acknowledge the financial support
of the National Research Council (KN) and Universities Space Research
Association (IMG, TJT, TY).}

\clearpage

% For the preprint version

\end{document}